\documentclass[aps,pra,groupaddress,twocolumn]{revtex4-1}
\usepackage{units}
\usepackage{amsmath}
\usepackage{amssymb}
\usepackage{graphicx}
\usepackage{bm}
\usepackage{multirow,color,relsize,ulem,microtype}

\newcommand{\be}{\begin{equation}}
\newcommand{\ee}{\end{equation}}

\newcommand{\re}[1]{\text{Re}[#1]}
\newcommand{\im}[1]{\text{Im}[#1]}

\newcommand{\omegafb}{\omega_{\scriptscriptstyle \text{FB}}}

\begin{document}

\title{Non-Hermitian lattices with a flat band and polynomial power increase [Invited]}

\author{Li Ge}
\email{li.ge@csi.cuny.edu}
\affiliation{\textls[-18]{Department of Engineering Science and Physics, College of Staten Island, CUNY, Staten Island, NY 10314, USA}}
\affiliation{The Graduate Center, CUNY, New York, NY 10016, USA}

\date{\today}

\begin{abstract}
In this work we first discuss systematically three general approaches to construct a non-Hermitian flat band, defined by its dispersionless real part. They resort to, respectively, spontaneous restoration of non-Hermitian particle-hole symmetry, a persisting flat band from the underlying Hermitian system, and a compact Wannier function that is an eigenstate of the entire system. For the last approach in particular, we show the simplest lattice structure where it can be applied, and we further identify a special case of such a flat band where every point in the Brillouin zone is an exceptional point of order 3. A localized excitation in this ``EP3 flat band" can display either a conserved power, quadratic power increase, or even quartic power increase, depending on whether the localized eigenstate or one of the two generalized eigenvectors is initially excited. Nevertheless, the asymptotic wave function in the long time limit is always given by the eigenstate, and in this case, the compact Wannier function or its superposition in two or more unit cells.\\

\end{abstract}

\maketitle

\section{Introduction}

A flat band, as the name suggests, is a dispersionless band which extends in the whole Brillouin zone. Systems that exhibit flat bands have attracted considerable interest in the past few years, including optical \cite{Apaja,Hyrkas} and photonic lattices \cite{Rechtsman,Vicencio,Mukherjee,Biondi}, graphene \cite{Kane,Guinea},  superconductors \cite{Simon,Kohler1,Kohler2,Imada}, fractional quantum Hall systems \cite{Tang,Neupert,Sarma} and exciton-polariton condensates \cite{Jacqmin,Baboux}. The flatness of the band leads to a zero group velocity, which has important implications on the dynamic and localization properties, including inverse Anderson transition \cite{Goda}, localization with unconventional critical exponents and multi-fractal behavior \cite{Chalker}, mobility edges with algebraic singularities~\cite{Bodyfelt}, and unusual scaling behaviors \cite{Flach,Leykam,Ge_AdP}.

For a Hermitian lattice, a completely flat band in the entire Brillouin zone is formed when there exists a Wannier function that is an eigenstate of the whole system. To understand this relation, we only need to resort to the definition of the Wannier function itself, which we denote by $W_n(x-ja)$ in one dimension (1D). Here $n$ is the band index, $a$ is the lattice constant, and $j$ is the unit cell index. The Bloch wave function with wave vector $k$ in the $n$th band can be written as
\be
\Psi_n(x;k) = \sum_j e^{ikaj}\,W_n(x-ja),
\ee
and it satisfies $H_0\Psi_n(x;k)=\omega_n(k)\Psi_n(x;k)$, where $H_0$ is the Hamiltonian of the entire system instead of the Bloch Hamiltonian $H(k)$ of the unit cell. Now if $H_0 W_n(x-ja) = \omega_w W_n(x-ja)$, i.e., there exists an Wannier function that is an eigenstate of the whole system with eigenvalue $\omega_w$, then we immediately find $\omega_n(k)=\omega_w$ which is $k$-independent.

\begin{figure}[hbt]
\centering
\includegraphics[width=\linewidth]{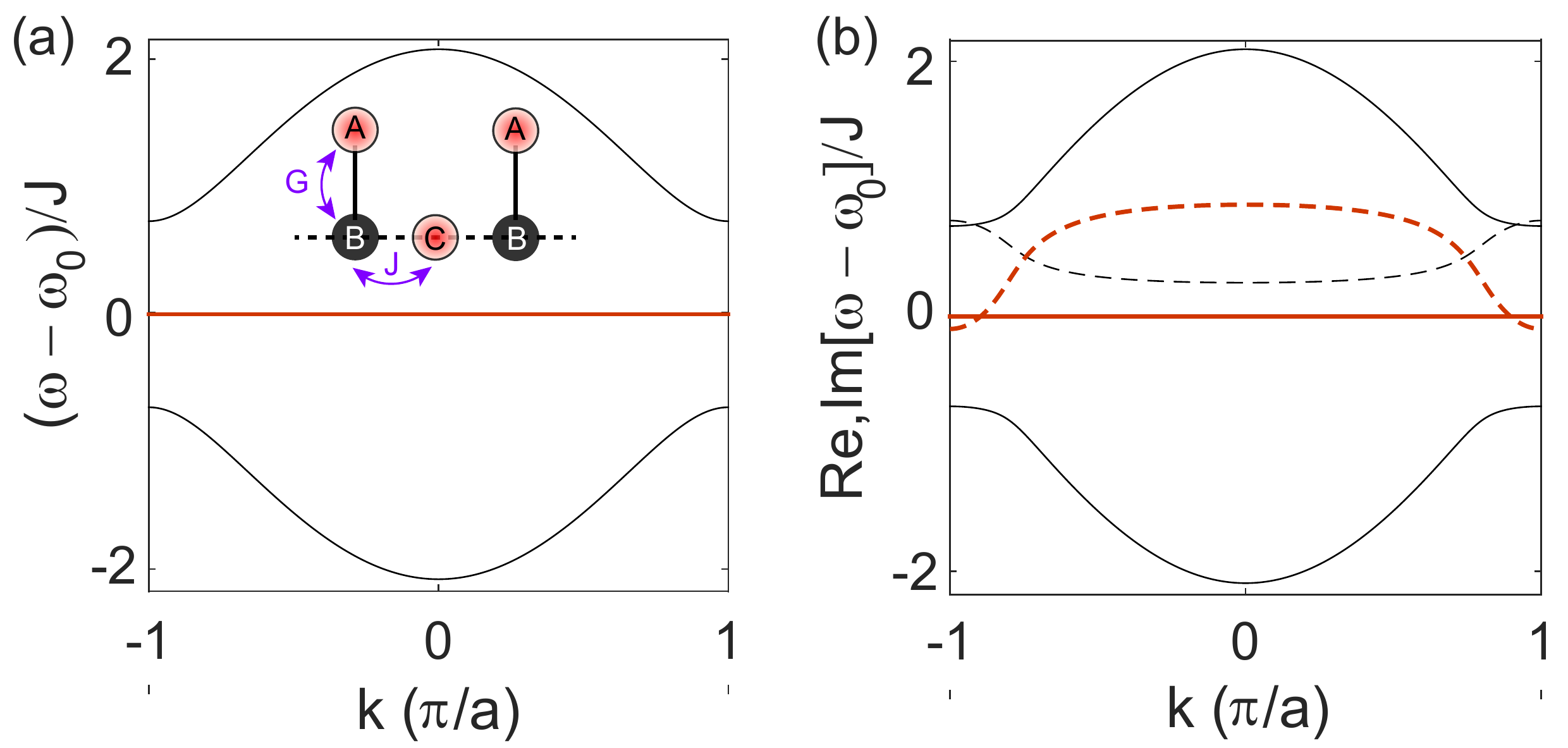}
\caption{(a) Band structure of a Hermitian Lieb lattice. The flat band is shown by the thick line. Inset: Schematic of the Lieb lattice, where $G$ (solid lines) is $3/4$ of $J$ (dashed lines). Partially transparent dots show the spatial profile of the compact Wannier function. (b) Same as (a) but with gain and loss modulation $\gamma_A = 1, \gamma_B = 0.5, \gamma_C = -0.1$. The dashed lines show the imaginary parts of the band structure, and those of the two dispersive bands are the same (given by the thin dashed line) due to the relation $\omega_l(k)=-\omega_m^*(k)\,(l\neq m)$ imposed by NHPH symmetry.}\label{fig:band}
\end{figure}

The simplest way to find such a Wannier function is in a frustrated lattice \cite{Miyahara}, where quantum tunnelings from the edges of the Wannier function to the neighboring unit cells interfere destructively and are completely cancelled, hence isolating the Wannier function from the rest of the lattice. Take the 1D Lieb lattice for example [see Fig.~\ref{fig:band}(a)]. It has an $L$-shaped unit cell, with decorated lattice sites (A) coupling to every other site (B) on the main lattice. In the tight-binding model it is captured by the Bloch Hamiltonian
\be
H(k) =
\begin{bmatrix}
\omega_A & G & 0 \\
G & \omega_B & J(1+e^{-ika}) \\
0 & J(1+e^{ika}) & \omega_C
\end{bmatrix},
\ee
where $G,J\in\mathbb{R}$ are the vertical and horizontal coupling coefficients. When there is no detuning between the A sites and the C sites on the main lattice (i.e., $\omega_A=\omega_C\equiv\omega_0$), a $V$-shaped Wannier function exists that spans two unit cells. This Wannier function has a nonzero amplitude only at the two A lattice sites and the C lattice site between them [see Fig.~\ref{fig:band}(a)], i.e.,
$[\psi^{(A)}_j,\psi^{(B)}_j,\psi^{(C)}_j,\psi^{(A)}_{j+1},\psi^{(B)}_{j+1},\psi^{(C)}_{j+1}]=[-J,0,G,-J,0,0]$,
and it is an eigenstate of the entire system with eigenvalue $\omega_w=\omega_0$. As a result, this Wannier function leads to a flat band at $\omegafb=\omega_0$.

Unlike condensed matter systems, the realization of flat bands in optics involves parallel waveguides or cavities that are coupled evanescently. As such, a new degree of freedom can be introduced to manipulate the forming and properties of flat bands in these systems, i.e., non-Hermiticity brought forth by optical gain and loss. As we shall see later, this additional tuning knob enables a more flexible control of band structures, with which a completely different approach can be employed to generate a flat band. Here it is worth pointing out that in a non-Hermitian system, the band structure is complex-valued in general and we define a flat band by requiring that its real part is $k$-independent.
The initial studies of flat band physics in a non-Hermitian system considered the effect of parity-time perturbations on a Hermitian flat band \cite{Flatband_PT,Chern,Molina}, where optical gain and loss are arranged in a judicious way to satisfy the parity-time symmetry \cite{NatPhoReview}. More recently, several studies probed the existence of non-Hermitian flat bands using either gain and loss modulations or complex-valued couplings \cite{FB_Konotop,FB_Hami,FB_Yidong,defectState}.

The goal of this article is to provide a unified view of how non-Hermitian flat bands, as defined above, can be constructed. More specifically, we discuss systematically three such approaches, namely, using spontaneous restoration of non-Hermitian particle-hole (NHPH) symmetry (``Approach 1" \cite{defectState}), a persisting flat band from the underlying Hermitian system (``Approach 2"), and a compact Wannier function that is an eigenstate of the entire system (``Approach 3" \cite{FB_Konotop,FB_Hami,FB_Yidong}).
For Approach 3 in particular, we give the simplest lattice structure where this approach can be applied, which contains only two lattice sites in a unit cell. We further identify a special case of such a flat band where every point in the Brillouin zone is an exceptional point (EP) of order 3 \cite{Flatband_PT,Graefe}, and a localized excitation in this “EP3 flat band” can display either a conserved power, quadratic power increase, or even quartic power increase, depending on whether the compact Wannier function or one of the two generalized eigenvectors is initially excited. Nevertheless, the asymptotic wave function in the long time limit is always given by an eigenstate, and in this case, the compact Wannier function or its superposition in two or more unit cells.

\section{Constructing non-Hermitian flat bands}

\subsection{Approach 1}
\label{sec:approach1}
Approach 1 mentioned above was first suggested in Ref.~\cite{defectState}, where the flat band is a result of spontaneously restored NHPH symmetry \cite{antiPT,Malzard,zeromodeLaser} for all modes in the flat band. NHPH symmetry requires that the Hamiltonian of the system anticommutes with an antilinear operator, i.e., $\{H,CK\}=0$, where $C$ is a linear operator and $K$ denotes the complex conjugation. This approach (and the NHPH symmetry) does not require a frustrated lattice, and the band structure in the Hermitian limit does not have a flat band; instead, it is universal for a system that consists of two sublattices with real-valued nearest neighbor coupling and identical $\omega_0$ for all lattice sites before the gain and loss modulation is introduced (i.e., $\omega_j\rightarrow\omega_j+i\gamma_j$ for site $j$) \cite{zeromodeLaser}. NHPH symmetry leads to a symmetric spectrum satisfying $[\omega_l(k)-\omega_0]=-[\omega_m(k)-\omega_0]^*$ where $l,m$ are two band indices, and we have $\re{\omega_l(k)}=\omega_0$ when $l=m$, which defines the symmetric phase of NHPH symmetry. As a result, a flat band at $\re{\omegafb}=\omega_0$ is formed. Note that since the system does not have a flat band in its Hermitian limit, the flat band is formed after the non-Hermitian perturbation collapses the real parts of an even number of previous dispersive bands \cite{Makris_prl08,Bo_Nature}. Therefore, this flat band actually contains at least two bands with degenerate real parts, which are distinguished only by their different imaginary parts.

\begin{figure}[b]
\centering
\includegraphics[clip,width=\linewidth]{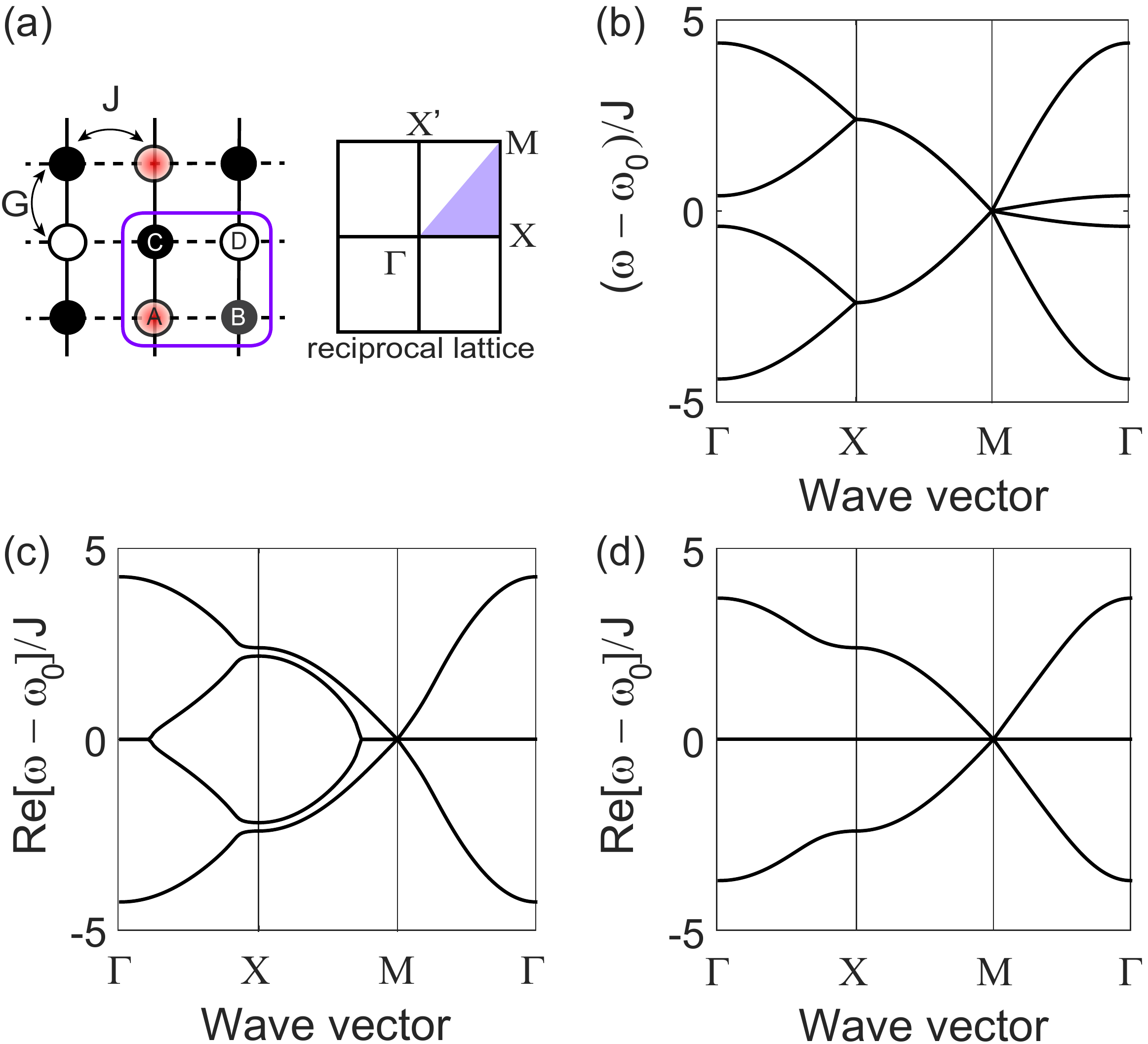}
\caption{Band structure of a 2D rectangular lattice with loss introduced to the A sites. (a) Schematics of the rectangular lattice and its reciprocal lattice. The unit cell is highlighted by the rounded box. $G=1.2J>0$. (b)-(d) Real part of the band structure when $\gamma_A/J=0,-2,-4.8$.}\label{fig:m2_2D}
\end{figure}

Here we exemplify a two-dimensional (2D) lattice with a flat band formed via this approach. We consider a rectangular lattice with a unit cell of four sites, with loss introduced only to the A site marked in Fig.~\ref{fig:m2_2D}(a). The lattice constants are denoted by $a$ and $b$ in the $x$ and $y$ directions. The Bloch Hamiltonian of this system can be written as
\be
H(k) =
\begin{bmatrix}
i\gamma_a & \tilde{J}(k_x) & \tilde{G}(k_y) & 0 \\
\tilde{J}^*(k_x) & 0 & 0 & \tilde{G}(k_y)\\
\tilde{G}^*(k_y) & 0 & 0 & \tilde{J}(k_x)\\
0 & \tilde{G}^*(k_y) & \tilde{J}^*(k_x) & 0
\end{bmatrix},
\ee
where $\tilde{J}(k_x)\equiv J(1+e^{-2ik_xa}), \tilde{G}(k_y)\equiv G(1+e^{-2ik_yb})$ and $J,G$ are again the nearest neighbor couplings in the $x$ and $y$ directions. It is easy to see that the system has NHPH symmetry: it consists of two sublattices formed by (A,D) and (B,C) lattice sites respectively, and all lattice sites have the same on-site energy ($\omega_0=0$) before the loss modulation is introduced.

\begin{figure}[t]
\centering
\includegraphics[clip,width=\linewidth]{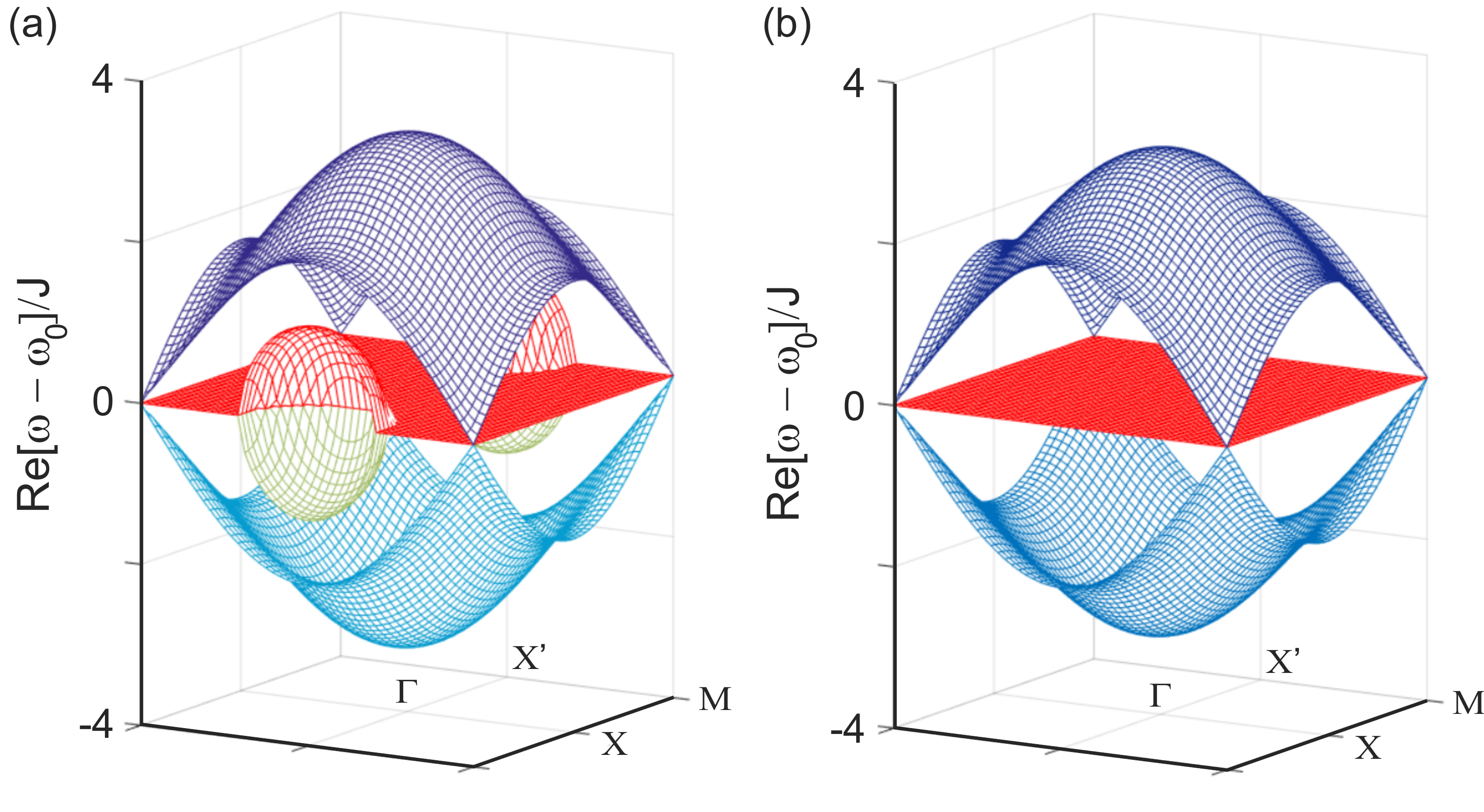}
\caption{Same as Fig.~\ref{fig:m2_2D} but plotted in three dimensions with $G=0.8J>0$ and $\gamma_A/J=-3.2$ (a), $-4$ (b).}\label{fig:m2_2D_2}
\end{figure}

The band structure of this rectangular lattice in the Hermitian limit ($\gamma_A\rightarrow0$) is shown in Fig.~\ref{fig:m2_2D}(b).
We note that the two sections between $X$ and $M$ in the first Brillouin zone are both doubly degenerate, because at $k_yb=\pm\pi/2$ the two rows of the unit cell are effectively decoupled [i.e., $\tilde{G}(k_y=\pm\pi/2b)=0$]. All four bands merge into a single point at the $M$ point, where $k_xa$ also becomes $\pm\pi/2$ and all couplings disappear effectively [i.e., $\tilde{G}(k_y=\pm\pi/2b)=\tilde{J}(k_x=\pm\pi/2a)=0$]. This observation at $M$ holds also in the non-Hermitian case (see Fig.~\ref{fig:m2_2D_2}).

\begin{figure}[t]
\centering
\includegraphics[clip,width=\linewidth]{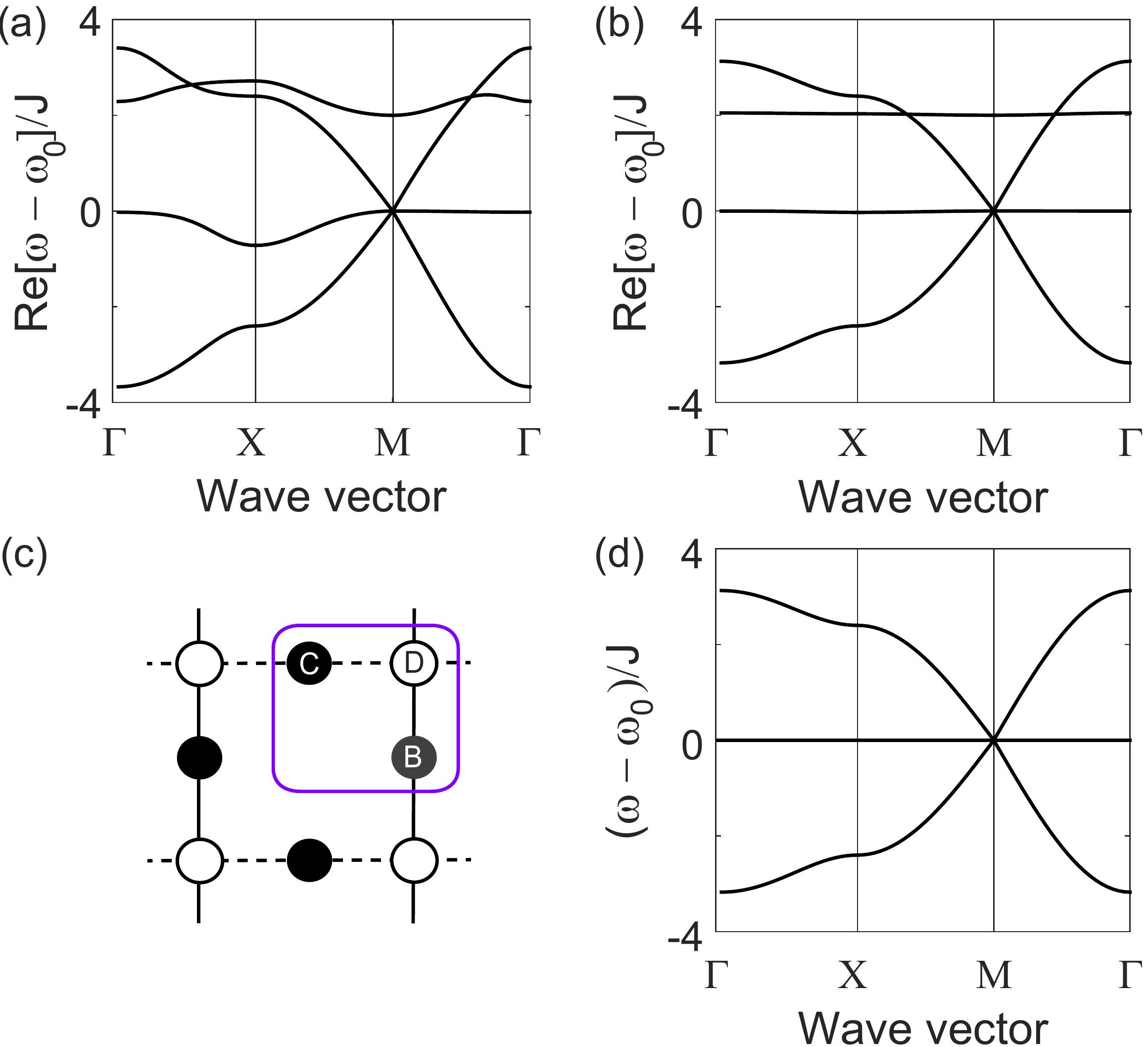}
\caption{Band structure of the 2D square lattice shown in Fig.~\ref{fig:m2_2D}(a) but with a detuning $\Delta_A=2J$ and $\gamma_A/J=-4.8$ (a), $-20$ (b). The effectively Hermitian Lieb lattice in (b) and its band structure are shown in (c) and (d).}\label{fig:m2_2D_3}
\end{figure}

As $|\gamma_A|$ increases, the central two bands start to collapse [Fig.~\ref{fig:m2_2D}(c)] and the non-Hermitian flat band is completed when $|\gamma_A|=4|G|$ [Fig.~\ref{fig:m2_2D}(d)]. Note that we have chosen $|G|>|J|$ here. If $|G|<|J|$ instead, the middle band seems to be flat when $|\gamma_A|=4|G|$ along the usual path $\Gamma$-$X$-$M$-$\Gamma$ in the first Brillouin zone, but in fact it is not flat yet near the $X'$ point ($k_x=0,k_yb=\pm\pi/2$) [Fig.~\ref{fig:m2_2D_2}(a)]. It is straightforward to show that in this case the flat band is completed when $|\gamma_A|=4|J|$ [Fig.~\ref{fig:m2_2D_2}(b)].

If we lift the NHPH symmetry of the system, e.g., by introducing a detuning $\Delta_A=2J$ at the A site in each unit cell, then a strictly flat band no longer exists [Fig.~\ref{fig:m2_2D_3}(a)]. We do note that a huge $\gamma_A$ in this case effectively decouples all the A sites from the rest of the lattice, and the latter actually form a Hermitian 2D Lieb lattice [Fig.~\ref{fig:m2_2D_3}(c)]. As a result, two approximately flat bands are formed in this limit [Fig.~\ref{fig:m2_2D_3}(b)], one with $\omega\approx\Delta_A+i\gamma_A$ from the isolated A sites and the other with $\omega\approx0$ originating from the Hermitian flat band of the 2D Lieb lattice [Fig.~\ref{fig:m2_2D_3}(d)] \cite{Vicencio,Mukherjee}; the latter is similar to its 1D counterpart shown in Fig.~\ref{fig:band}(a), and the former is a trivial example of Approach 3 where the Wannier function is on a single A site, not because of frustration but due to loss-induced isolation.

\subsection{Approach 2}
\label{sec:approach2}
Another, more intuitive approach (``Approach 2") to construct a non-Hermitian flat band is to start from a Hermitian system that has a flat band. The aim is then to maintain this Hermitian flat band (at least its real part) after the introduction of gain and loss. For the 1D Lieb lattice mentioned in the introduction, if the same $\gamma$ is added to the non-dark sites (A and C) of the Wannier function, the flat band persists trivially with $\omegafb=\omega_0+i\gamma$, whether or not $\omega_B$ becomes complex. In this case the Wannier function is still isolated from the rest of the lattice, and it is a special case of Approach 3 we will discuss in detail later. We also mention that the Hermitian flat band studied in Ref. \cite{Flatband_PT} persists with a parity-time symmetric perturbation \cite{El-Ganainy_OL07,Moiseyev,Musslimani_prl08,Makris_prl08,Guo,conservation,EP9,Ruter}
largely due to this reason. Surprisingly, even when an arbitrary gain and loss configuration is imposed in the unit cell of the Lieb lattice shown in Fig.~\ref{fig:band}(a), the real part of the central band is still flat if $\omega_B=\omega_0$, with its imaginary part now exhibiting a non-trivial $k$-dependence [see Fig.~\ref{fig:band}(b)]. This observation indicates that there ceases to exist a Wannier function that is an eigenstate of the whole system, and the mechanism that leads to the flat band is different from the underlying Hermitian case.

\begin{figure}[t]
\centering
\includegraphics[width=\linewidth]{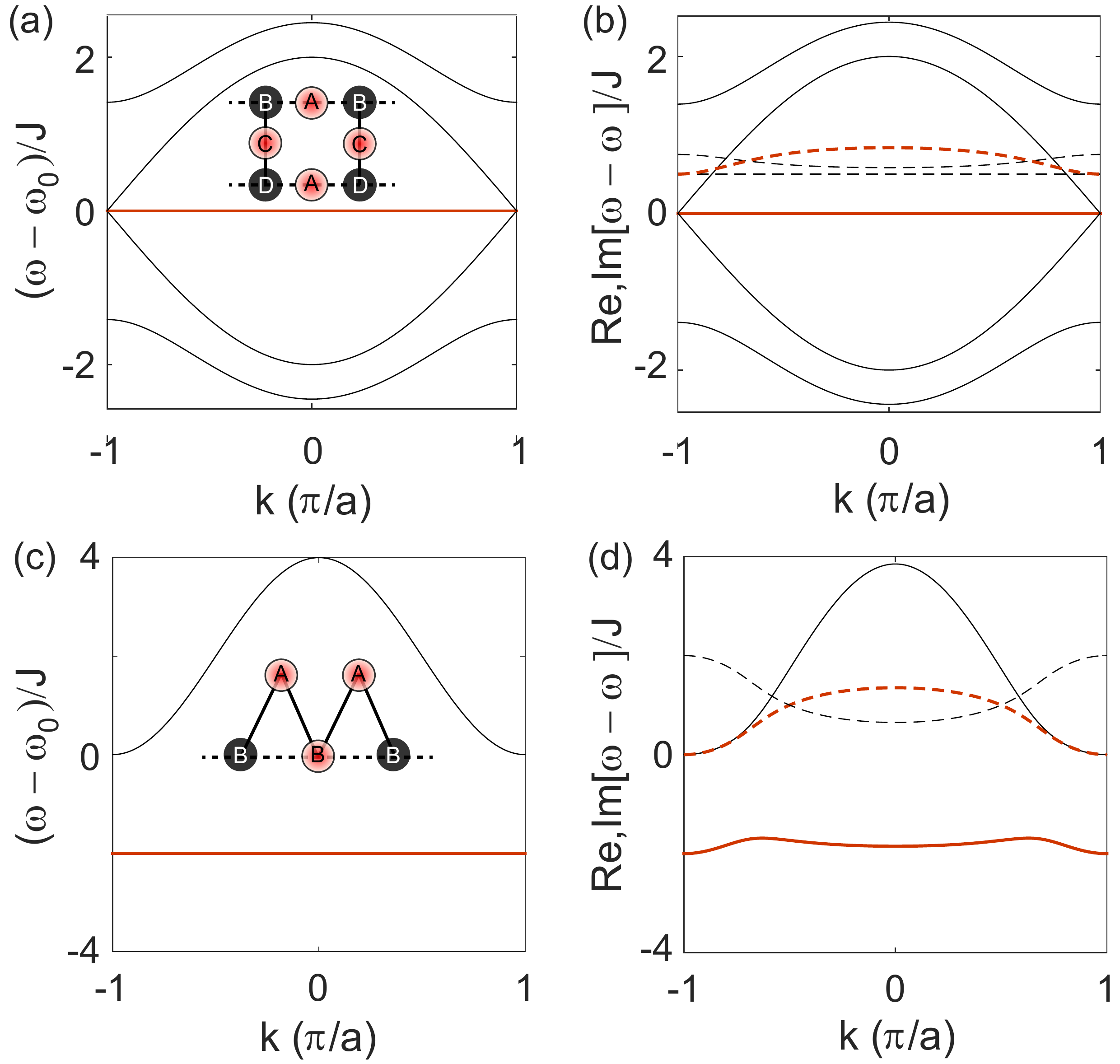}
\caption{Persisting and non-persisting Hermitian flat bands. (a) Band structure of a Hermitian quasi-1D edge-centered square lattice. Inset: $G$ (solid lines) $=J$ (dashed lines). Partially transparent dots show the spatial profile of the compact Wannier function. (b) Same as (a) but with $\gamma_{A,B,D,E}=0.5,\gamma_C=1$. Again the imaginary parts (dashed lines) of the dispersive bands are identical pairwise due to NHPH symmetry. (c),(d) Same as (a),(b) but for a saw lattice. $\gamma_A=0,\gamma_B=2$. Inset in (c): $G$ (solid lines) $=\sqrt{2}J$ (dashed lines).
}\label{fig:band2}
\end{figure}

One way to understand this behavior is again using NHPH symmetry. This system has NHPH symmetry even with an arbitrary gain and loss modulation, and the flat band modes stay in the symmetric phase of the NHPH symmetry where $\re{\omega}=\omega_0$. However, this consideration does not tell us immediately why these modes do not undergo a spontaneous NHPH symmetry breaking when the gain and loss modulation is strong, which would lift the flatness of the band.

As it turns out, this persisting flat band can also be understood as the result of (i) a well-known mathematical theorem: one root of a cubic equation with real coefficients is always real; and (ii) the underlying Hermitian flat band is at the identical on-site energy, i.e., $\omegafb=\omega_0$.
To see how these two conditions enforce a non-Hermitian flat band, we write the eigenvalues of the periodic system as $\omega(k)\equiv\omega_0+iu(k)$, where $u(k)$ is complex in general. We then find that the equation for $u$ is a cubic equation $u^3+bu^2+cu+d=0$ with only real coefficients
$b = \sum_p \gamma_p$, $c = -G^2 -2(1+\cos ka)J^2+ \sum_{q>p}\gamma_p\gamma_q$, and $d =(G^2-\gamma_A\gamma_B)\gamma_C + 2(1+\cos ka)\gamma_AJ^2$,
where the summations run through the three lattice sites in a unit cell. Therefore, $u(k)$ always has a real root according to (i), which indicates that $\omega(k)-\omega_0$ always has an imaginary root, i.e., a flat band at $\re{\omega(k)}=\omega_0$ with system- and $k$-dependent $\im{\omega(k)}=u(k)$. It is then clear that this flat band is independent of the strength of the gain and loss modulation.

\begin{figure}[t]
\centering
\includegraphics[clip,width=\linewidth]{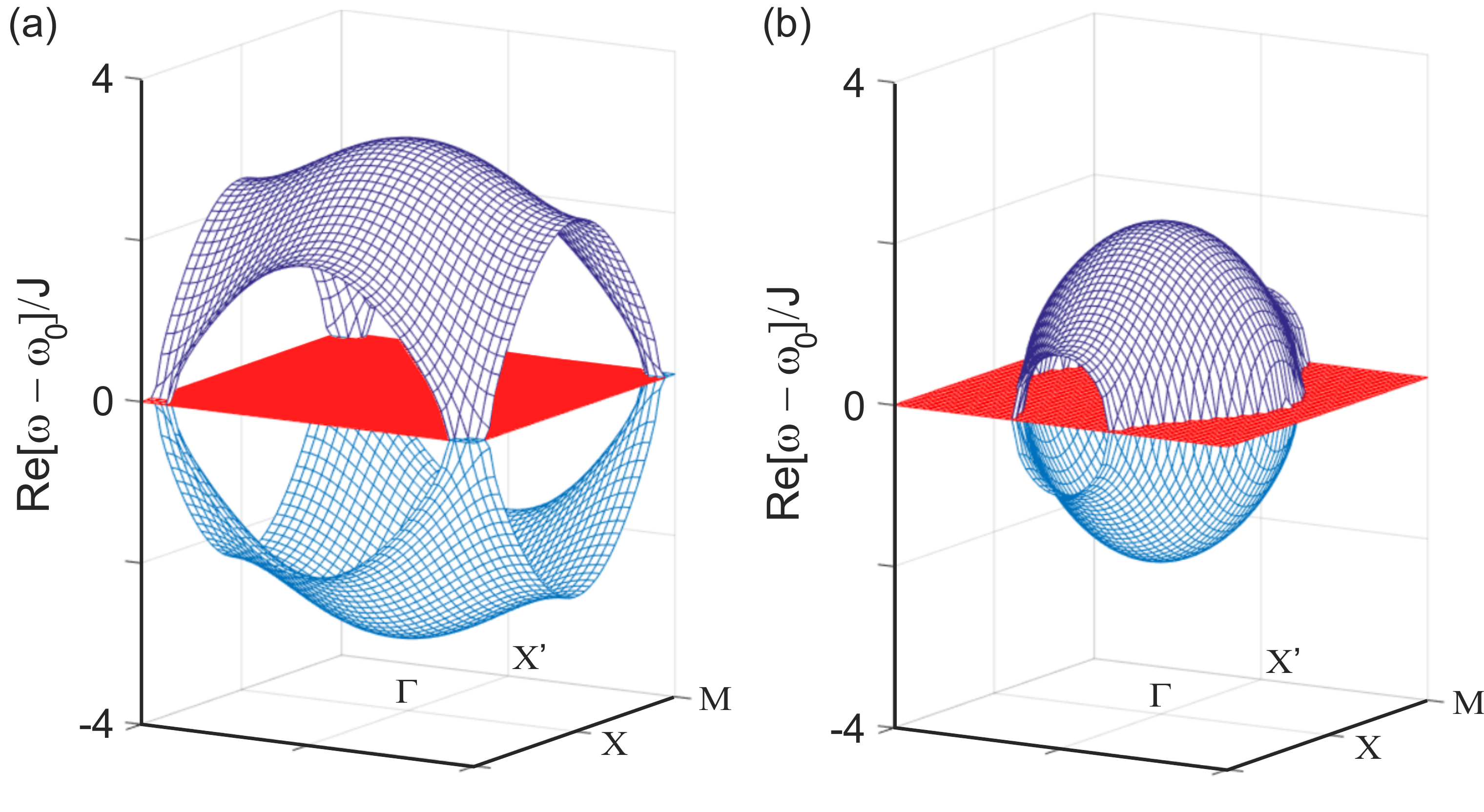}
\caption{Persisting flat band in a 2D non-Hermitian Lieb lattice with $G=1.2J>0$. $\gamma_{B,C,D}=-0.78,-0.71,0.32$ in (a) and $-2.1,-0.3,3.3$ in (b).}\label{fig:m2_2D_4}
\end{figure}

In fact, this property holds for many frustrated lattices with an odd number ($q$) of lattice sites in a unit cell but not for those with an even $q$, as we exemplify in Figs.~\ref{fig:band2}(b) and \ref{fig:band2}(d) with $q=5$ and 2. This approach is not limited to 1D and quasi-1D lattices; it also applies, for example, to the 2D Lieb lattice shown in Fig.~\ref{fig:m2_2D_3}(c) with $q=3$. Two sets of randomly generated gain and loss modulations in a unit cell are shown in Fig.~\ref{fig:m2_2D_4}.

\subsection{Approach 3}
\label{sec:approach3}
The third approach (``Approach 3") to construct a non-Hermitian flat band is to construct a localized Wannier function that is an eigenstate of the whole system. Similar to the Hermitian case mentioned in the introduction, this approach can be applied to a frustrated lattice and the resulting flat band has a $k$-independent imaginary part as well. One example of this approach was given in Ref.~\cite{FB_Hami} for a triangle lattice, and two other examples are given in Ref.~\cite{FB_Yidong}. The difference in these two similar studies is that the former considered a lattice that does not have a flat band without gain or loss, while the latter used lattices that do have a flat band in the Hermitian limit. In this sense, the latter is similar to Approach 2 in spirit, with a non-Hermitian perturbation that still makes the Wannier function an eigenstate of the whole system.

\begin{figure}[t]
\centering
\includegraphics[width=\linewidth]{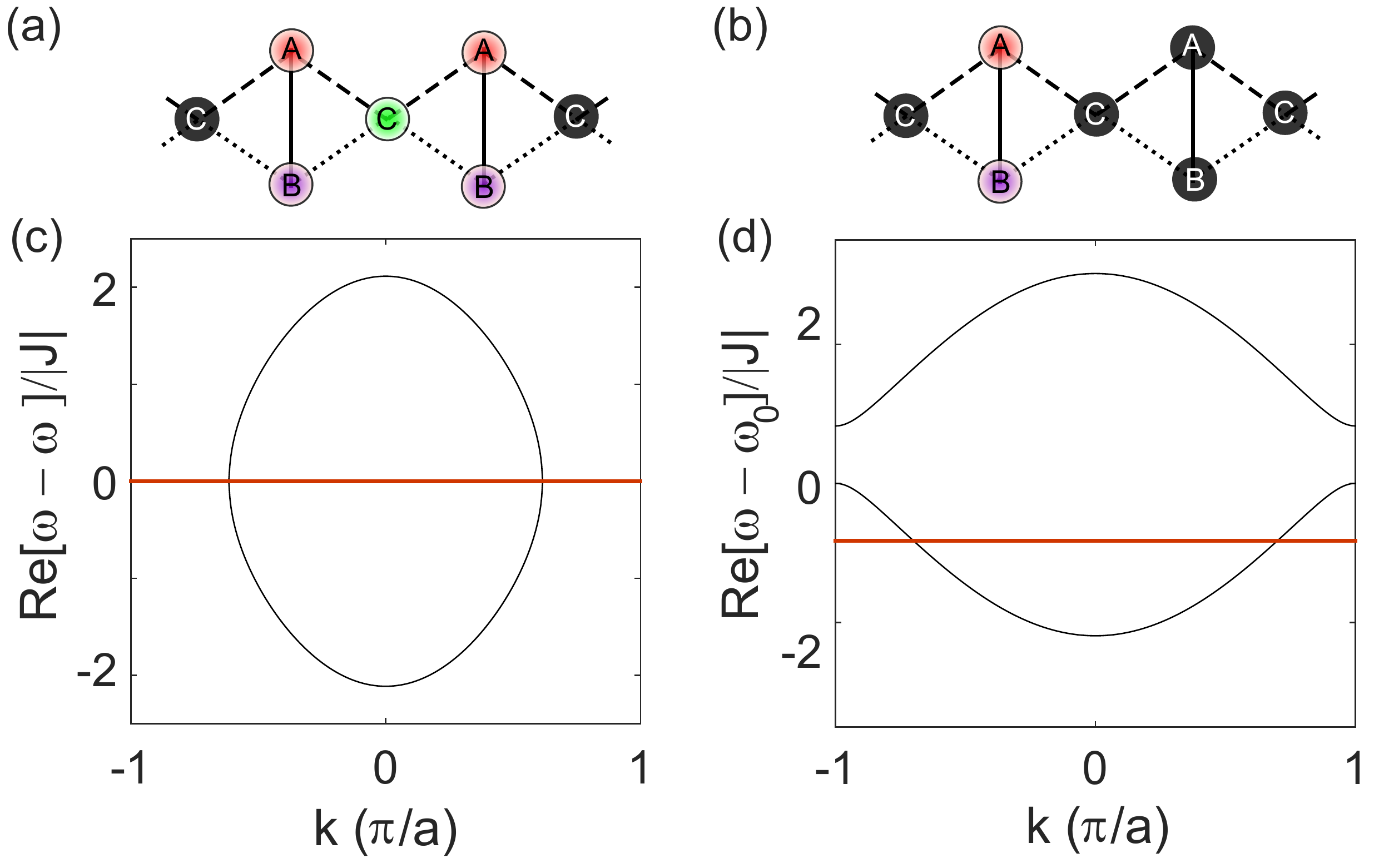}
\caption{(a,b) Two compact Wannier functions (partially transparent dots) for the cross-stitched lattice shown. Couplings are represented by dashed lines ($J$), dash-dotted lines ($J^*$), and solid lines ($G$). Gain ($i\gamma$) and loss ($-i\gamma$) are introduced to B and A sites, respectively. Here $G=|J|$, $\text{Arg}[J]=0.3\equiv\theta/2$, and $\gamma=-G/\sin\theta$ in (a) and $-G\sin\theta$ in (b). (c,d) Band structure of this cross-stitched lattice for the values of $\gamma$ in (a) and (b). The flat bands result from the compact Wannier functions in (a) and (b).
}\label{fig:cross}
\end{figure}

We should mention that two non-Hermitian flat bands similar in construction were also found numerically in Ref.~\cite{FB_Konotop}, but the authors there did not discuss the origin of the flatness or the existence of a compact Wannier function in the non-Hermitian case for either flat band. We point out here that the Wannier function for these two flat bands span one and two unit cells, respectively. More specifically, the model considered in Ref.~\cite{FB_Konotop} can be written as
\be
H(k) =
\begin{bmatrix}
-i\gamma & G & J(1+e^{-ika})\\
G & i\gamma & J^*(1+e^{-ika}) \\
J(1+e^{ika}) & J^*(1+e^{ika}) & 0
\end{bmatrix}\label{eq:cross}
\ee
and its cross-stitched lattice is shown schematically in Figs.~\ref{fig:cross}(a) and \ref{fig:cross}(b). Here we have shifted the spectrum (such that $\omega_0=0$) to simplify the notation. Denoting $\text{Arg}[J]\equiv\theta/2$, a flat band is formed when $\gamma=-G/\sin\theta$ or $-G\sin\theta$, where $\omegafb=0$ and $-G\cos\theta$, respectively [see Figs.~\ref{fig:cross}(c) and \ref{fig:cross}(d)]. The compact Wannier functions in these two cases are shown in Figs.~\ref{fig:cross}(a) and \ref{fig:cross}(b), and they are given by $[\psi^{(A)}_j,\psi^{(B)}_j,\psi^{(C)}_j]=[-1,e^{i\theta},0]$ in a single unit cell and by $[\psi^{(A)}_j,\psi^{(B)}_j,\psi^{(C)}_j,\psi^{(A)}_{j+1},\psi^{(B)}_{j+1},\psi^{(C)}_{j+1}]=[-1,e^{i\theta},iG\cos\theta/(J^*\sin\theta),-1,e^{i\theta},0]$ in two unit cells, respectively.

All the three references mentioned above considered a unit cell with three lattice sites, which however is not the simplest lattice structure that can have a Wannier function with the aforementioned property; the simplest one is the saw lattice shown in Fig.~\ref{fig:band2}(c), now with a complex coupling $G=J\sqrt{2+i(\gamma_A-\gamma_B)/J}$:
\be
H(k) =
\begin{bmatrix}
i\gamma_A & G(1+e^{-ika})\\
G(1+e^{ika}) & i\gamma_B + 2J\cos ka
\end{bmatrix}.\label{eq:saw}
\ee
Interestingly, the non-Hermitian perturbations $\gamma_{A,B}$ and the resulting $G$ \textit{do not} change the real part of the band structure at all; the imaginary parts of the flat band and dispersive band are both flat and equal to $\gamma_B$ and $\gamma_A$, respectively:
\be
\omegafb = -2J + i\gamma_B,\quad \omega_{\scriptscriptstyle \text{D}} = 2J(1+\cos ka) + i\gamma_A.\label{eq:saw_flat}
\ee
The Wannier function that leads to the non-Hermitian flat band has the same form as in the Hermitian case, i.e., $[\psi^{(A)}_j,\psi^{(B)}_j,\psi^{(A)}_{j+1},\psi^{(B)}_{j+1}]=[-1,t,-1,0]$ in two unit cells.

\section{The presence of EPs and polynomial power increase}

An EP is a non-Hermitian degenerate point where not only the eigenvalues but also the wave functions of two or more eigenstates coalesce \cite{EP1,EP2,EP3,EPMVB,EP4,EP5,EP6,EP8,EP10}. EPs are found in both Refs.~\cite{FB_Hami,FB_Yidong}, and the formation of the non-Hermitian flat band is attributed to these EPs, to some extent \cite{FB_Hami}. However, from our discussion of Approach 3 above and especially the example given by Eq.~(\ref{eq:saw}), it is clear that this approach does not rely on the existence of an EP.

When all system parameters are fixed and the wave vector $k$ is the only variable, Ref.~\cite{FB_Hami} found that a non-Hermitian flat band can have two EPs at two different values of $k$. Ref.~\cite{FB_Yidong} identified a scenario that every possible state in the flat band corresponds to an EP of order 2. Here we first point out that in fact a higher order EP can be found for every wave vector in the Brillouin zone, without increasing the size of the unit cell (i.e., 3 lattice sites). In addition, we unveil polynomial power increase in such an ``EP3 flat band," which can display either quadratic or quartic behaviors.

To exemplify this EP3 flat band, we turn to the lattice described by Eq.~(\ref{eq:cross}) and require $\theta=(p+0.5)\pi\,(p\in Z)$ and $\gamma=-G\sin\theta=\pm G$. Below we take $p$ to be an odd integer, and the case with an even $p$ is the same with a few sign changes. We now have $\gamma=G$, $J^*=iJ$, and the Bloch Hamiltonian becomes
\be
H(k) =
\begin{bmatrix}
-iG & G & J(1+e^{-ika})\\
G & iG & iJ(1+e^{-ika}) \\
J(1+e^{ika}) & iJ(1+e^{ika}) & 0
\end{bmatrix}.\label{eq:cross2}
\ee
We immediately find that the first two rows (columns) of $H(k)$ in this case differ by a factor of $i$, and hence the Bloch Hamiltonian has a zero eigenvalue, independent of the value of $k$; this is exactly the flat band $\omegafb=-G\cos\theta$ mentioned in the previous section. More importantly, the characteristic polynomial of $H(k)$ is simply $\omega^3=0$, again independent of $k$. Therefore, every state in this flat band corresponds to an EP of order 3 \cite{Graefe,Flatband_PT}, including the collapsed and identical Wannier function for all values of $k$ that is given by $[\psi^{(A)}_j,\psi^{(B)}_j,\psi^{(C)}_j]=[1,i,0]$.

\begin{figure*}[t]
\centering
\includegraphics[width=\linewidth]{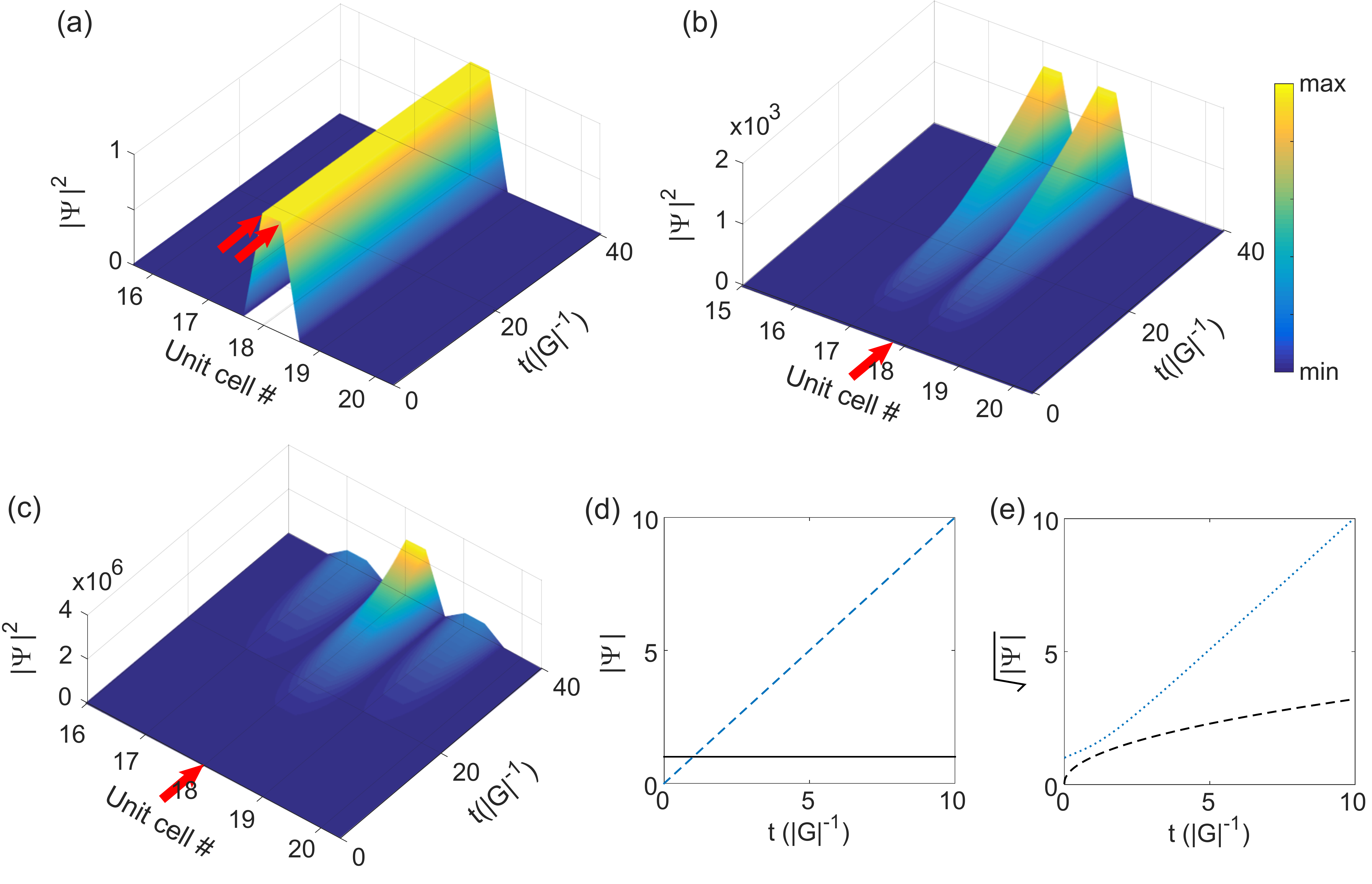}
\caption{Polynomial power dependence for a localized initial excitation in an EP3 flat band. The excitation of (a) a Wannier function in a single unit cell, (b) a C site, and (c) an A site are shown by the red arrows. Note the different scales of the vertical axis in these three panels, even thought the initial amplitudes of the excitation all equal 1. (d) Fixed amplitude $|\Psi^{(C)}|=1$ at the excitation site in (b) (solid line) and linearly increasing amplitude $|\Psi^{(A)}|$ next to it (dashed line). (e) Quadratically increasing amplitude $|\Psi^{(A)}|$ at the excitation site in (c) (dotted line) and linearly increasing amplitude $|\Psi^{(C)}|$ next to it (dashed line).
}\label{fig:dyna}
\end{figure*}

An initial excitation of this Wannier function or its superposition in more than one unit cell leads to a conserved energy in this EP3 flat band, since the corresponding eigenvalue is real [see Fig.~\ref{fig:dyna}(a)]. For a more general excitation, we resort to the notion of generalized eigenvectors \cite{linearAlgebra} to investigate the dynamics in the system. Let $\Psi_0$ and $\omega_0$ be an triply coalesced eigenstate and eigenvalue of the system Hamiltonian $H$. We then define the first generalized eigenvector $\Psi_1$ by
\be
[H-\omega_0 \mathbf{1}]\Psi_1 = \Psi_0, \label{eq:gv1}
\ee
where $\mathbf{1}$ is the identity matrix. A particular consequence of Eq.~(\ref{eq:gv1}) is
\be
e^{-iHt}\Psi_1 = e^{-i\omega_0t}(\Psi_1 -it \Psi_0),\label{eq:quadratic}
\ee
i.e., an initial excitation of $\Psi_1$ will display a quadratic power increase in the asymptotic limit $t\rightarrow\infty$ when $\omega_0$ is real [see Fig.~\ref{fig:dyna}(b)], and the asymptotic wave function is given by the eigenvector $\Psi_0$ itself. The latter observation holds even when $\omega_0$ is complex.
In the EP3 flat band discussed above, we found that $\Psi_1$ can be chosen simply as $\Psi_1=[\psi^{(A)}_j,\psi^{(B)}_j,\psi^{(C)}_j]=[0,0,1]$ in a single unit cell, and the corresponding eigenstate is an equal superposition of the Wannier functions of two neighboring unit cells, i.e., $\Psi_0=[\psi^{(A)}_j,\psi^{(B)}_j,\psi^{(A)}_{j+1},\psi^{(B)}_{j+1}]=[J,iJ,J,iJ]$; the only exception takes place when the $j$th unit cell is the last one, where $\Psi_0=[\psi^{(A)}_N,\psi^{(B)}_N]=[J,iJ]$. To simplify the notations, we have dropped from these expressions the lattice sites with a zero amplitude.

Similarly, we define the second generalized eigenvector $\Psi_2$ by
\be
[H-\omega_0 \mathbf{1}]\Psi_2 = \Psi_1,\label{eq:gv2}
\ee
which indicates
\be
e^{-iHt}\Psi_2 = e^{-i\omega_0t}\left(\Psi_2 -it \Psi_1 - \frac{t^2}{2}\Psi_0\right).\label{eq:quartic}
\ee
Therefore, we can expect a quartic ($t^4$) power increase as $t\rightarrow\infty$, if $\omega_0$ is real and $\Psi_2$ is excited initially [see fig.~\ref{fig:dyna}(c)]. It is important to note that for an extended system, $\Psi_1$ in Eq.~(\ref{eq:gv2}) does not need to be identical to that defined in Eq.~(\ref{eq:gv1}); it can be a superposition of the latter in more than one unit cell, with the additional freedom of superposing the eigenstates in different unit cells as well. For example, we find that the tightest $\Psi_2$ is given by $\psi^{(A)}_j$=1 in the EP3 flat band discussed above, and the corresponding $\Psi_1$ is given by $[\psi^{(C)}_j,\psi^{(A)}_{j+1},\psi^{(B)}_{j+1},\psi^{(C)}_{j+1}]=[J,-iG,G,J]$, which also leads to a different $\Psi_0$ from that shown in Fig.~\ref{fig:dyna}(b), now given by $[\psi^{(A)}_{j-1},\psi^{(B)}_{j-1},\psi^{(A)}_j,\psi^{(B)}_j,\psi^{(A)}_{j+1},\psi^{(B)}_{j+1}]=[J^2,iJ^2,2J^2,2iJ^2,J^2,iJ^2]$; this is the asymptotic wave function shown in Fig.~\ref{fig:dyna}(c), which agrees with the analytical result given by Eq.~(\ref{eq:quartic}).
Note that we can also choose the second generalized eigenvector as $\Psi'_2=\psi^{(B)}_j=-i$, which leads to the same $\Psi_{1,0}$ as the $\Psi_2$ specified above; hence the quadratic and quartic power terms are cancelled for an initial excitation of the superposition $\Psi_2-\Psi'_2=[\psi^{(A)}_j,\psi^{(B)}_j]=[1,i]$, which is exactly the eigenstate itself in a unit cell.

\section{Discussion and Conclusion}

\begin{figure}[t]
\centering
\includegraphics[width=0.8\linewidth]{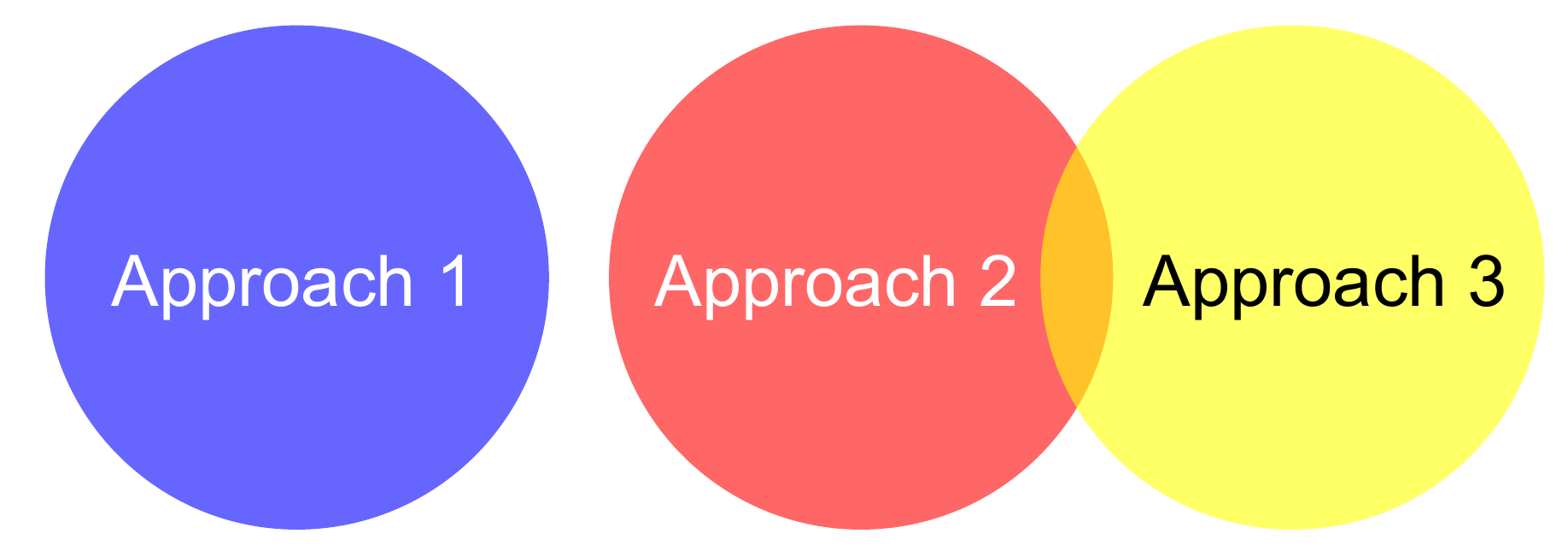}
\caption{Relation between the three general approaches to construct a non-Hermitian flat band.}\label{fig:overlap}
\end{figure}

In conclusion, we have discussed systematically three approaches to achieve a non-Hermitian flat band. Their relation is summarized in Fig.~\ref{fig:overlap}. Approach 2 and 3 overlap when the Hermitian Wannier function remains an eigenstate of the system, either with just the introduction gain and loss modulation (as already mentioned at the beginning of Sec.~\ref{sec:approach2}) or with complexified coupling(s) too (as the last case in Sec.~\ref{sec:approach3} shows). Approach 1 and 2, however, do not overlap, because in the Hermitian limit one of them has a flat band (Approach 2) but not the other (Approach 1). Similarly, Approach 1 and 3 do not overlap, because a non-Hermitian flat band constructed by Approach 1 is due to the spontaneously restored NHPH symmetry and has a $k$-dependent imaginary part of the flat band. In contrast, the flat band constructed in Approach 3 has a $k$-independent imaginary part instead.

Even though Approach 1 and 2 do not overlap, they still have a delicate relationship from the perspective of NHPH symmetry. When the couplings in Approach 2 are allowed to be complexified, the system itself does not need to have NHPH symmetry as the last case in Sec.~\ref{sec:approach3} shows. Otherwise, NHPH symmetry of the system is required in this approach as in Approach 1: For all the cases with a persisting Hermitian flat band considered in Sec.~\ref{sec:approach2} [i.e., Figs.~\ref{fig:band}(b), \ref{fig:band2}(b), \ref{fig:m2_2D_4}], there exist two sublattices where all couplings are real and take place only between sites on different sublattices, and hence the system has NHPH symmetry \cite{zeromodeLaser}; for the case with a non-persisting Hermitian flat band considered in Fig. \ref{fig:band2}(d), two neighboring B sites couple and hence the system does not have NHPH symmetry. Another case is the one shown in Fig.~\ref{fig:cross}(d): by taking $J$ to be real (and hence $\theta=0$), a Hermitian flat band lays at $\omegafb=-G$. Now even though each unit cell of this lattice has three lattice sites, similar to the Lieb lattice in Fig.~\ref{fig:band}(b), it does not have the two sublattices required by NHPH symmetry. As a result, this flat band does not persist with the introduction of an arbitrary gain and loss modulation (not shown), if we keep all couplings real-valued.

Although only the last approach, i.e., constructing a Wannier function that is an eigenstate of the entire system, leads to a flat imaginary part of the dispersion relation as well, the existence of localized defect states with a slight perturbation does not rely on this additional property \cite{defectState}. Since all implementations of flat bands in a photonic structure has defects, whether using laser-written waveguides [3-5] or microcavities fabricated by various etching methods [6], the effect of the dispersive imaginary part does not affect the non-Hermitian flat bands from revealing this key manifestation of their Hermitian counterpart. If for some applications a vanished imaginary part of the entire flat band is preferred, Approach III should be adopted. For example, the two flat bands shown in Fig.~\ref{fig:cross} are both real valued, and the one given by Eq.~(\ref{eq:saw_flat}) can be made real by having $\gamma_B=0$. For the EP3 flat band we have exemplified using Approach III, a quartic power increase is expected unless the initial excitation is exactly an eigenstate or the first generalized eigenvector. This behavior is very different from previous found quadratic power increase due to an EP of order 2 \cite{Longhi_PRA}.

For simplicity, we have discussed mainly quasi-1D lattices, but the results presented here can be easily generalized to two-dimensional lattices as we have exemplified in Sec.~\ref{sec:approach1} and \ref{sec:approach2}. In fact, the additional dimension does not need to be a spatial dimension; it can come from the internal energy structure of the lattice sites, when we consider, for example, cold atoms in an optical lattice \cite{ZHE}.
Experimentally preparing a non-Hermitian system at an EP remains to be very difficult, but recent successes of demonstrating EP-based sensing schemes \cite{EPsensing1,EPsensing2} have proved that such a challenge can be overcome with finely tuned optical systems.\\

\noindent \textbf{Funding.} National Science Foundation (NSF) (DMR-1506987). \\

\noindent \textbf{Acknowledgement.} The author acknowledges support from NSF.

\bibliographystyle{apsrev}

\end{document}